\documentclass[prd,twocolumn,showpacs,preprintnumbers,aps,nofootinbib]{revtex4-1}

\usepackage{graphicx}% Include figure files
\usepackage{dcolumn}% Align table columns on decimal point
\usepackage{bm}% bold math
\usepackage{amsmath,amssymb,amsfonts}
\usepackage{latexsym}

\usepackage{color}
\def\l{\lambda}
\def\bmtoeg{\mathcal{B}(\mu\to e \gamma)}
\def\bttomg{\mathcal{B}(\tau\to \mu \gamma)}
\def\mtoeg{\mu\to e \gamma}
\def\ttomg{\tau\to \mu \gamma}
\def\mntoen{\mu N\to e N}

\def\rtm{\rho_{\tau\mu}}
\def\rmt{\rho_{\mu\tau}}
\def\rte{\rho_{\tau e}}
\def\ret{\rho_{e\tau}}
\def\rtata{\rho_{\tau\tau}}

\def\rtt{\rho_{tt}}

\begin{document}

\title{\boldmath
Charged lepton  flavor violation in light of Muon $g-2$
}

\author{Wei-Shu Hou and Girish Kumar}
\affiliation{
Department of Physics, National Taiwan University, Taipei 10617, Taiwan}
% \affiliation{ }
\bigskip

% \date{\today}

\begin{abstract} 
The recent confirmation of the muon $g-2$ anomaly by the Fermilab g-2 experiment
may harbinger a new era in $\mu$ and $\tau$ physics. In the context of
general two Higgs doublet model, the discrepancy can be explained via
one-loop exchange of  sub-TeV exotic scalar and pseudoscalars,
namely $H$ and $A$, that have  flavor changing neutral couplings
$\rho_{\tau\mu}$ and $\rho_{\mu\tau}$ at $\sim 20$ times the
usual tau Yukawa coupling, $\lambda_\tau$.
Taking $\rho_{\ell\ell^\prime}\sim \lambda_{ \rm min(\ell, \ell^\prime)}$,
we show that the above solution to muon $g-2$ then predicts enhanced rates of
various charged lepton flavor violating processes, which should be accessible
at upcoming experiments. We cover muon related processes such as
$\mu \to e \gamma$, $\mu \to eee$ and $\mu N \to e N$, and $\tau$ decays
$\tau \to \mu \gamma$ and $\tau \to \mu\mu\mu$. A similar one-loop diagram with
$\rho_{e\tau}= \rho_{\tau e} = {\cal O}(\lambda_e)$ induces $\mu \to e\gamma$,
bringing the rate right into the sensitivity of the MEG~II experiment.
The $\mu e\gamma$ dipole can be probed further by $\mu \to 3e$ and
$\mu N \to eN$. With its promised sensitivity range and ability to
use different nuclei, the $\mu N \to eN$ conversion experiments can
not only make discovery, but access the extra diagonal quark Yukawa
couplings $\rho_{qq}$. For the $\tau$ lepton, we find that
$\tau \to \mu\gamma$ would probe $\rho_{\tau\tau}$ down
to $\lambda_\tau$ or lower, while $\tau \to 3\mu$ would
probe $\rho_{\mu\mu}$ to ${\cal O}(\lambda_{\mu})$.
\end{abstract}

\maketitle

%-----------------------------------------------------------------------------------------------------------------------------------
%	Introduction
%-----------------------------------------------------------------------------------------------------------------------------------

\section{Introduction}

In 1948, Schwinger presented his result~\cite{Schwinger:1948iu}
for the ``anomalous'' magnetic moment of the electron, 
$a_e \equiv (g_e - 2)/2\simeq \alpha/2\pi$. 
The observable has played one of the most important roles in establishing particle physics: 
consistency between prediction and experiment has established QED as 
the most accurate fundamental theory of {\it Nature} known to humankind.
In the last two decades, with experiments able to perform ever
precise measurements to expose the tiniest deviations, muon $g-2$ 
has become a flagship observable in the search for New Physics (NP), 
or physics beyond the Standard Model (SM).
Recent developments suggest a possible revival of muon (and tau) physics,
as we illustrate.

The Fermilab Muon g-2 experiment~\cite{Muong-2:2021ojo} reported recently 
its measurement of the $g-2$ of the muon, 
%$a_\mu\equiv (g-2)/2$, 
confirming the previous result at Brookhaven~\cite{Muong-2:2006rrc}. 
Combining the two measurements~\cite{Muong-2:2021ojo} 
gives
\begin{align}
a_\mu^{\rm Exp} = 116 592 061(41) \times 10^{-11}.
\label{g-2_2021}
\end{align}
Compared with the community-wide
theory consensus~\cite{Aoyama:2020ynm,
Aoyama:2012wk,Aoyama:2019ryr,Czarnecki:2002nt,Gnendiger:2013pva,Davier:2017zfy,Davier:2010nc,
Keshavarzi:2018mgv,Colangelo:2018mtw,Hoferichter:2019mqg,Davier:2019can,Keshavarzi:2019abf,
Kurz:2014wya,Melnikov:2003xd,Masjuan:2017tvw,Colangelo:2017fiz,Hoferichter:2018kwz,
Gerardin:2019vio,Bijnens:2019ghy,Colangelo:2019uex,Blum:2019ugy,Colangelo:2014qya}, 
$a_\mu^{\rm SM} = 116 591 810(43) \times 10^{-11}$,
the difference %between %Eq.~\eqref{g-2_2021} and Eq.~\eqref{a_mu_sm}, 
\begin{align}
\Delta a_\mu = a_\mu^{\rm Exp} - a_\mu^{\rm SM} =  (251 \pm 59) \times 10^{-11},
\label{Del-a_mu}
\end{align}
has a significance of $4.2\sigma$~\cite{Muong-2:2021ojo}.

This large discrepancy has certainly attracted much attention. 
We refer the reader to Ref.~\cite{Athron:2021iuf} for 
a recent review of popular NP models that provide solutions to
the muon $g-2$ anomaly, whereas a slightly dated review 
can be found in Ref.~\cite{Lindner:2016bgg}.
One of the desired ingredients to ease the NP explanation 
%of Eq.~\eqref{Del-a_mu} 
is chiral enhancement. 
In this regard, the general two-Higgs doublet model (g2HDM), 
sometimes referred to as 2HDM Type-III~\cite{Hou:1991un},
is one of the simplest extensions of SM which can do the job.

\begin{figure}[b]
\center
\includegraphics[width=0.27 \textwidth]{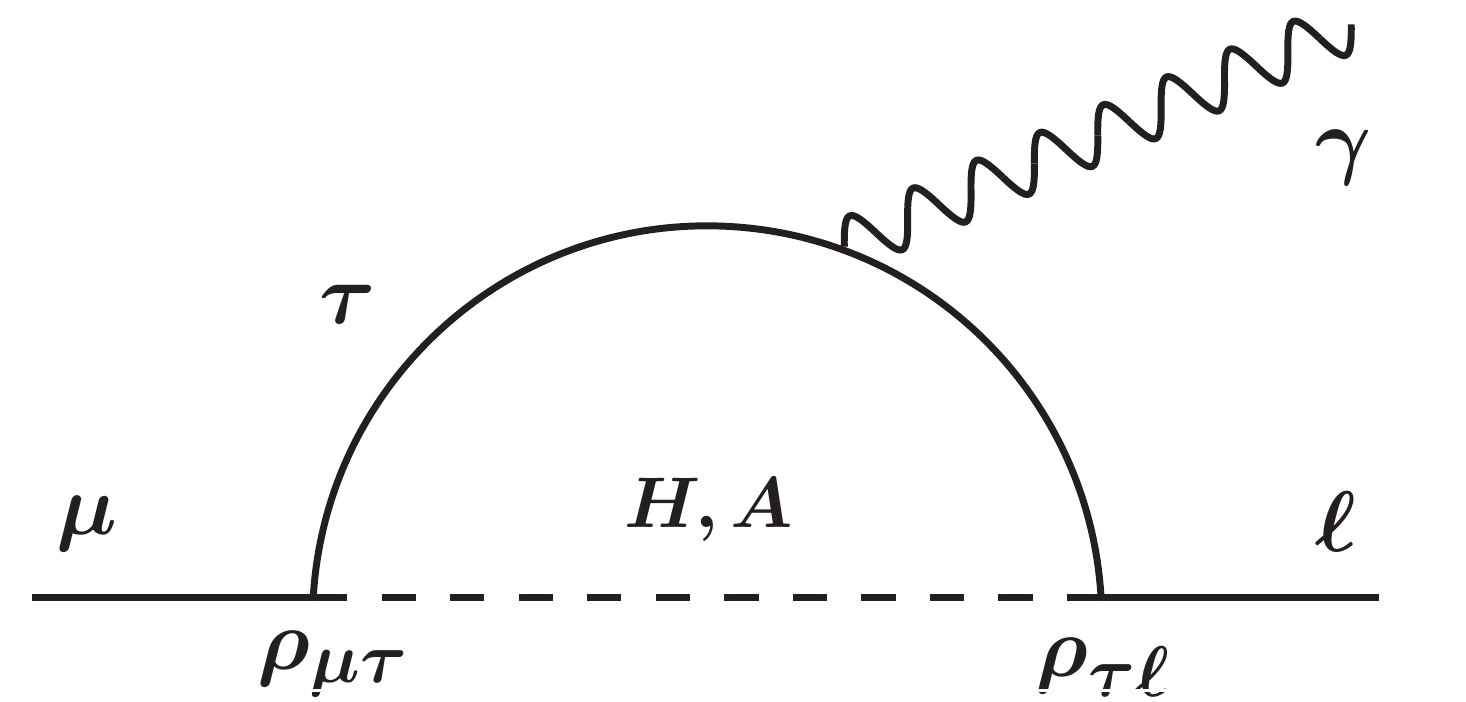}
\caption{One-loop diagram for muon $g-2$ (with $\ell=\mu$) and $\mu\to e\gamma$ (with $\ell = e$).}
\label{fig: one-loop}
\end{figure}

The usual 2HDM  Type-II impose 
a $Z_2$ symmetry to remove flavor violation from the
Lagrangian~\cite{Branco:2011iw}.
In contrast, g2HDM does not adopt this symmetry, 
but {one} keeps all possible {Yukawa}
coupling terms between fermions and both scalar doublets. 
Thus, there exists {\it extra} flavor changing neutral couplings (FCNC) 
such as $\rtm$ ($\rmt$) that relate to lepton flavor violation (LFV) effects 
in the $32$ sector.  
As illustrated in Fig.~\ref{fig: one-loop}, these couplings 
can give rise to muon $g-2$ via the one-loop diagram, 
with tau and heavy neutral scalars $H$ and $A$ in the loop. 
The diagram enjoys $m_\tau/m_\mu$ chiral enhancement
and can explain Eq.~\eqref{Del-a_mu} for  $\rtm=\rmt$ at ${\cal O}(20\l_\tau)$, 
where $\lambda_\tau = \sqrt2 m_\tau/v$, % is the tau Yukawa coupling in SM,
with sub-TeV but nondegenerate $m_H$ and $m_A$, 
as we showed recently in Ref.~\cite{Hou:2021sfl}.

The strength of $\rtm$ and $\rmt$ causes concern about 
LFV decay of the observed SM Higgs boson $h$, 
where~\cite{CMS:2021rsq}
\begin{equation}\label{eq: h2taumu}
	{\cal B}(h\to\tau\mu) < 0.15 \%,
\end{equation}
and is also warranted for other off-diagonal extra {Yukawa} couplings
such as $t \to ch$~\cite{Chen:2013qta, CMS:2021bdg}.
But this can be easily tackled by noting~\cite{Chen:2013qta} that 
the strength of the $h \bar f_i f_j$ vertex ($i\ne j$) 
is proportional to $\rho_{ij} c_\gamma$,
where $c_\gamma \equiv \cos\gamma$ is the $h$--$H$ mixing angle. 
{\it Alignment}, that $c_\gamma$ seems quite small, emerged 
from detailed studies of $h$ boson properties after its 
discovery~\cite{CMS:2012qbp, ATLAS:2012yve, Khachatryan:2016vau}:
the $h$ boson resembles rather closely the SM Higgs.
The smallness of $c_\gamma$ suppresses~\cite{Hou:2017hiw} $h$ boson FCNC 
naturally, and constraints such as Eq.~\eqref{eq: h2taumu} can be evaded. 

We will take the alignment limit of $c_\gamma \to 0$ in this work.
This means the absence of flavor violating interactions of the $h$ boson, 
while for the exotic $H$ and $A$ bosons they appear at full strength
 ({i.e.} $\sin\gamma =1$).
The neutral exotic {Yukawa} couplings 
%that contain the FCNCs of the g2HDM Lagrangian then 
simplify to~\cite{Davidson:2005cw},
\begin{align}
 %\supset
  & \frac{1}{\sqrt{2}} \sum_{f = u, d, \ell} 
 \bar f_{i}  \,\rho^f_{ij} \left[ H
    + i\,{\rm sgn}(Q_f)  A \right]  R\, f_{j} 
 +{\rm h.c.},
\label{eq:Yuk}
\end{align}
where $R=(1+\gamma_5)/2$ is the right-handed projection.
The $\rho_{ij}$ couplings are in general complex,
where the phases can provide new sources of $CP$ violation 
that can drive electroweak baryogengesis 
(EWBG)~\cite{Chiang:2016vgf, Fuyuto:2017ewj, Fuyuto:2019svr, Hou:2020tnc, Modak:2018csw}.

In this paper, we seek to explore charged LFV phenomena 
involving $\mu$ and $\tau$, with solving the muon  $g-2$ anomaly in the backdrop. 
This is especially salient in g2HDM,
where the one-loop solution %to muon $g-2$ 
requires nonzero $\tau\mu$ FCNC, 
therefore correlates directly with LFV.
Previously, we highlighted~\cite{Hou:2020tgl, Hou:2020itz} 
the two-loop mechanism, with $\rho_{tt}\simeq {\cal O}(\l_t)$ 
as the main driver of LFV. 
This was based on identifying two experimentally viable textures, viz.
$\rho_{3i} \lesssim \l_3$ $(i\ne 1)$ and $\rho_{1j} \lesssim \l_1$.
The one-loop solution to muon  $g-2$ anomaly 
requires large deviation from the first condition, 
which in turn suppresses $\rho_{tt}$, hence the two-loop mechanism. 
But the second condition is not affected. 
With $\rho_{\tau\mu}\rho_{\mu\tau}$ 
fixed by the one-loop solution to muon  $g-2$,
we continue to adhere to 
\begin{equation}\label{eq: l_1j}
	\rho_{1j} \lesssim {\cal O}(\l_1),
\end{equation}
as a working assumption. 
These modifications change the conclusion from Ref.~\cite{Hou:2020itz} drastically. 
While we suggested that MEG~II~\cite{MEGII:2018kmf} 
would run into ``diminished return'' in its probe of $\mtoeg$, 
the one-loop solution to muon $g-2$, together with 
$\rte\lesssim \lambda_e$, Eq.~\eqref{eq: l_1j}, 
puts MEG II at the {\it cusp} of discovery, as we will show.
 
Many works have discussed charged LFV processes in g2HDM
previously, in the context of the muon $g-2$ 
anomaly~\cite{Crivellin:2013wna, Omura:2015nja, Omura:2015xcg, Wang:2016rvz, Primulando:2016eod, Crivellin:2019dun}.
In particular, Ref.~\cite{Omura:2015xcg} comes closest to this work.
Let us therefore point out and contrast what is new in the present work. 
First, most other works were written in a time when 
there was a hint for $h\to\tau\mu$~\cite{CMS:2015qee}, 
and therefore necessarily required finite --- and highly tuned ---
values of $c_\gamma$. The hint quickly evaporated, however,
and the latest CMS bound of Eq.~\eqref{eq: h2taumu} implies
%that accommodating even as small as 
$c_\gamma \lesssim 10^{-2}$~\cite{Hou:2021sfl}.
%values which are consistent with muon  $g-2$ is difficult;
%
Second, we shall highlight $\mntoen$ 
as the ultimate probe of LFV in g2HDM. 
Both $\mu e \gamma$ dipole and $\mu eqq$ contact terms,
as well as their interference, play important roles,
and can be used to infer the sign of mass splitting, $\Delta m = m_A - m_H$, 
which is important for the explanation of muon $g-2$ in g2HDM.
In a similar vein, Ref.~\cite{Omura:2015xcg} considered
only tree level contributions to $\mu \to 3e$ and $\tau \to 3\mu$.
Lastly, we avoid using any {\it cosmetic} cancellation mechanism
between one- and two-loop contributions to $\ell \to \ell^\prime \gamma$
for sake of enlarging parameter space.
We will lay out the reasons for this choice
when we discuss the two-loop mechanism.

This paper is organized as follows.
In Sec.~II  we discuss $\mtoeg$ in g2HDM and highlight 
the interplay of LFV in $32$ and $13$  sectors.
With the former couplings fixed by Eq.~\eqref{Del-a_mu},
the couplings associated with the latter, 
at ${\cal O}(\lambda_e)$ or less,
are shown to be well within the sensitivity of MEG~II to probe. 
Implications for $\mu\to 3 e$ are discussed.
Turning to $\mu N \to eN$ in Sec.~III, 
we discuss both dipole and contact contributions.  
In Sec.~IV we discuss $\ttomg$ and $\tau\to3\mu$
and their experimental prospects.
Finally, we discuss in Sec.~V other constraints 
and implications, % for electron $g-2$, 
and offer our conclusion.

\section{${\bm \mtoeg}$}\label{sec: II}

The leading contribution to $\mtoeg$ in g2HDM arises also
through one-loop diagrams, as shown in Fig.~\ref{fig: one-loop}.
Only $\tau$ is shown in the loop, as diagrams with
muon and electron are chiral-suppressed
%They suffers from small lepton
%mass-suppression and therefore will be sufficiently small.
and ignored. LFV in $\mtoeg$ arises from $\mu\tau$ and $\tau e$ FCNC. 

Defining the relevant effective Lagrangian 
as~\cite{Cirigliano:2009bz}
\begin{equation}
	\begin{aligned}
\mathcal{L}_{\mathrm{eff}}^{\rm \mu e\gamma} =
 & m_{\mu} \bigl(C_{T}^{R}\, \bar{e} \sigma_{\alpha\beta} L \mu
                         + C_{T}^{L}\, \bar{e} \sigma_{\alpha\beta} R \mu\bigr)
                         F^{\alpha\beta} \label{eq:Leff-dipole},
\end{aligned}
\end{equation}
the $\mtoeg$ branching fraction can be written
in terms of the Wilson coefficients $C_T^{L, R}$,
\begin{equation}
	{\cal B}(\mu\to e\gamma) =
	 \frac{48\pi^2}{G_F^2}\left(|C_T^{L}|^2+|C_T^{R}|^2\right).
\end{equation}
The diagram in Fig.~\ref{fig: one-loop} gives %for $C_T^R$, 
%TODO check expressions
\begin{equation}\begin{aligned} \label{eq:1-loop}
	C_{T}^{R}|_{\phi = H, A} \simeq 
(\pm) 
       \frac{ \rho_{ \mu \tau}^{\ast }\rho_{\tau e}^{\ast} }{64 \pi^{2}}
       \frac{em_\tau}{m_\mu m^2_\phi} \bigl[\log(m_\phi^2/m_\tau^2)- {3}/{2}\bigr],
\end{aligned}\end{equation}
where $+(-)$ sign is for the $H(A)$ contribution,
and some minor term has been dropped. 
To obtain $C_T^{L}$, one replaces $\rho_{ij}^\ast \to \rho_{ji}$.
% in Eq.~\eqref{eq:1-loop}.
The diagram with $H^+$ and neutrino in the loop 
is suppressed by neutrino mass.
% In our numerical analysis, however, we do \tcg{keep all the terms}.
To further simplify our numerics, 
we treat $\rho_{ij}$ as real\footnote{If one takes $\rho_{ij}$ to be complex,
the imaginary part of couplings will
induce new contributions to 
lepton electric dipole moment. A detailed study of
such CP violating observables in g2HDM, and their testability at experiments, has been
carried-out in Ref~\cite{Hou:2021zqq}.}
and, unless specified otherwise, take $\rho_{ij}= \rho_{ji}$.

   %TODO muon g-2
Before turning to numerical results for $\mtoeg$, let us 
quickly recall the muon  $g-2$ solution in g2HDM~\cite{Hou:2021sfl}. 
This will provide a constraint on $\rtm$ and
help define benchmark masses for heavy scalars.

The one-loop formula for muon  $g-2$ is
easily obtained from Eq.~\eqref{eq:1-loop} 
by change of label from ``e" to ``$\mu$",
\begin{equation}\label{eq: a_mu}
	\Delta a_\mu  = \frac{4 m_\mu^2}{e}
	\sum_{\phi=H, A} {\rm Re}\left[ C_T^{R}\right]_\phi\,,
\end{equation}
%
%As seen from Eq.~\eqref{eq:1-loop}, 
where $H $ and $A $ effects are opposite in sign.
Thus, to have finite $a_\mu$, $H$ and $A$ cannot be degenerate, 
or $\Delta m = m_A - m_H \ne 0$.
Choosing the sign of $\Delta m$, {i.e.} to have $H$ or $A$ 
lighter, is a matter of taste.
To keep consistency with our previous work~\cite{Hou:2021sfl}, 
we take $H$ lighter and fixed at $m_H = 300$ GeV, 
and take $\Delta m = 40$ and 200~GeV.
The close to degenerate $H$ and $A$ case
requires $\rtm\simeq 30 \l_\tau$ for $1\sigma$ solution of muon $g-2$.  
For the large splitting case, the effect of $A$ is damped with $H$ dominant, 
and a smaller $\rtm\simeq 20 \l_\tau$ suffices~\cite{Hou:2021sfl}.

\begin{figure*}[t]
\center
\includegraphics[width=.35\textwidth]{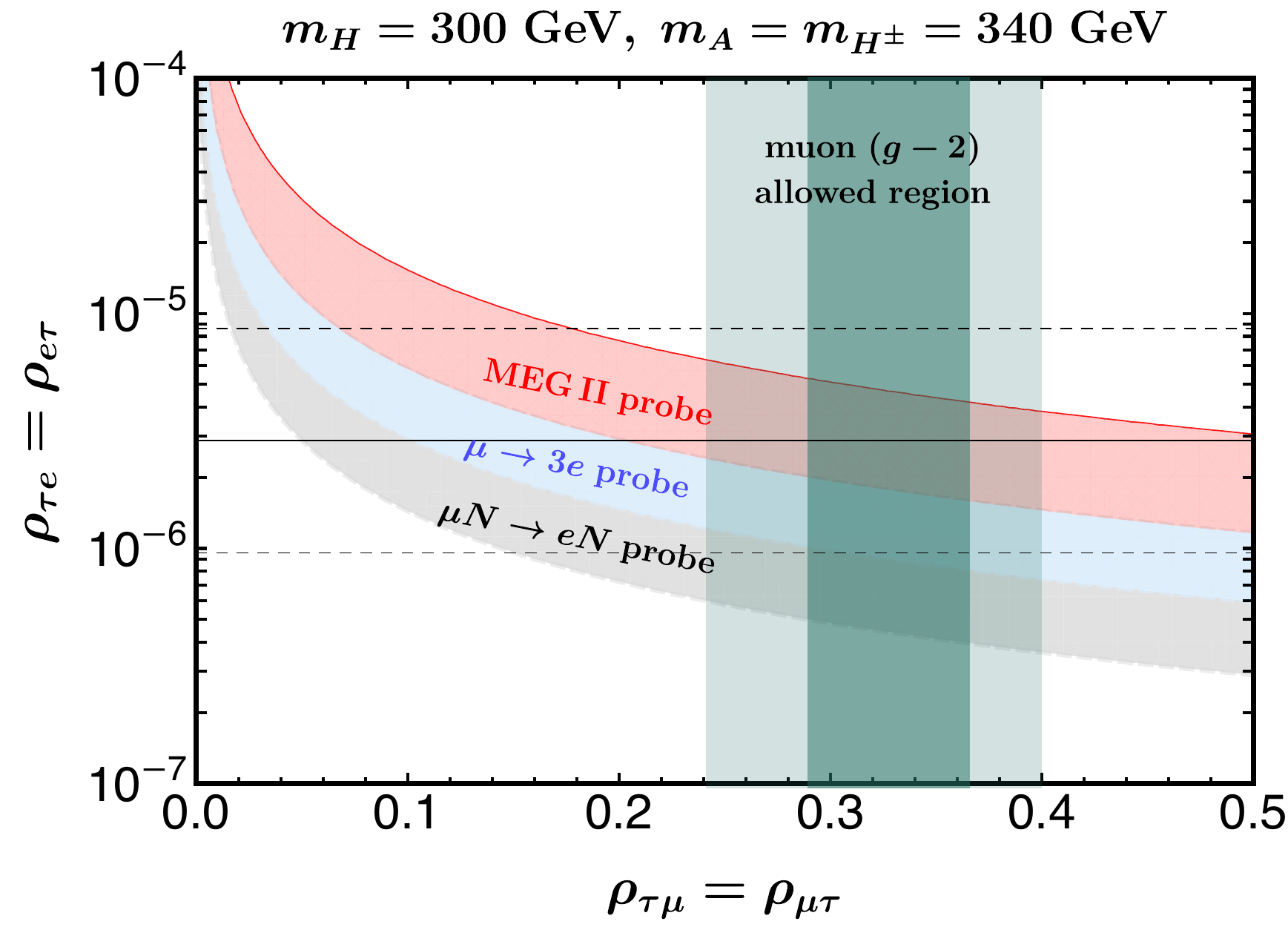}
\hskip 0.3cm
\includegraphics[width=.35\textwidth]{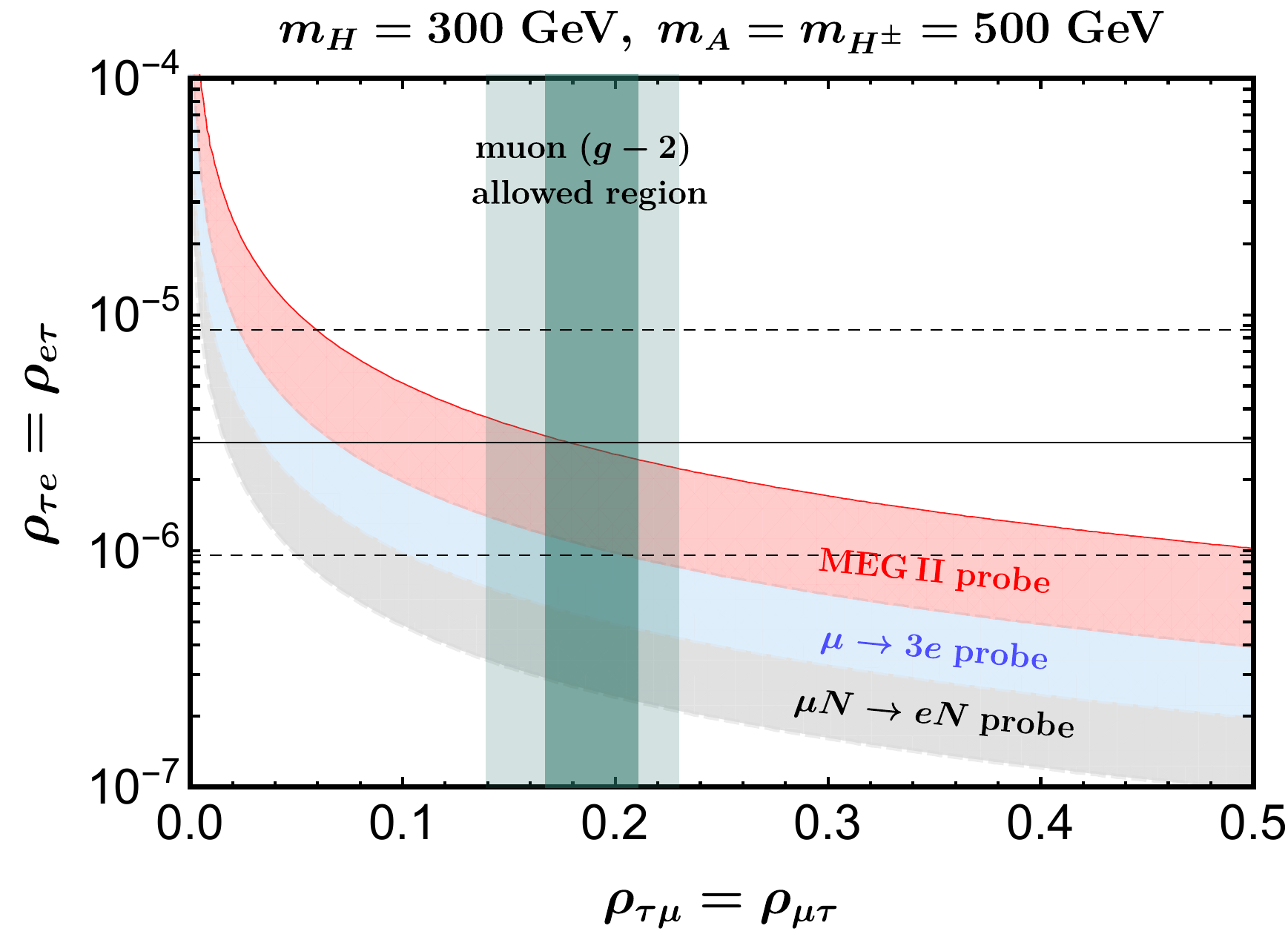}
\caption{
Parameter space in $\rte$--$\rtm$ plane which will be probed 
by three muon flavor violating processes. The dark (light) 
green shaded region is consistent with muon  $g-2$ within 
$1\sigma$ ($2\sigma$).  See text for details.}
\label{fig:mufv}
\end{figure*}

Returning to $\mtoeg$, utilizing Eq.~\eqref{eq:1-loop},
we plot in Fig.~\ref{fig:mufv} the region (red shaded) in 
the $\rtm$--$\rte$ plane to be probed by MEG~II~\cite{MEGII:2018kmf},
for $\Delta m = 40$ (200) GeV in the left (right) plot. 
The upper boundary corresponds to the MEG bound of 
$\bmtoeg_{\rm MEG}< 4.2 \times 10^{-13}$~\cite{MEG:2016leq},
and the lower boundary is the projected 
reach of MEG~II~\cite{MEGII:2018kmf}, 
$\bmtoeg_{\rm MEG ~II}< 6 \times 10^{-14}$. 
The parameter space consistent within $1\sigma $ ($2\sigma$) range
of muon $g-2$, Eq.~\eqref{Del-a_mu}, 
is highlighted as the dark (light) green shaded area.  
As we continue to advocate Eq.~\eqref{eq: l_1j} i.e. $\rte\lesssim {\cal O}(\l_e)$ 
as a natural choice for the electron-related off-diagonal coupling, 
we illustrate $\rte =\l_e$ and $\rte = 3\l_e, \, \l_e/3$ 
by horizontal solid and dashed lines, respectively.
It is intriguing that $\rte = \l_e$ sits right 
in the middle of the region that MEG~II would probe. 
One also sees that $\rte \gtrsim 3\l_e$ is
already ruled out by MEG, while $\rte\lesssim \l_e/3$ 
or smaller will fall short of the MEG~II range.
However, if  $m_A$ is large compared to $m_H$, 
as shown in Fig.~\ref{fig:mufv}(right),
then MEG II can probe down to $\rte\simeq \l_e/3$. 
 
Our working assumption of Eq.~\eqref{eq: l_1j} therefore suggests 
that MEG~II might well make a discovery.

   %TODO 2-loop
{\it \textbf{Two-loop contributions}.---} \
It is well-known that two-loop contributions,
the so-called Barr-Zee diagrams, can dominate over
one-loop contributions in certain parameter space of g2HDM.
The corresponding formulae and loop functions for $\mtoeg$
were originally calculated in Ref.~\cite{Chang:1993kw}, 
where it was shown that large extra top {Yukawa} coupling can drive
these contributions well above the one-loop diagram just discussed. 
However, these Barr-Zee diagrams
depend on $\rho_{\mu e}$ and diagonal $\rho_{ff}$
($f=t$ being the dominant loop contribution),
which do not play any direct role in our NP interpretation
of muon $g-2$. Therefore, these diagrams provide constraint
on the product of $\rho_{\mu e}\rtt$.
A more detailed phenomenological exploration of these
contributions can be
found in our previous work~\cite{Hou:2020itz}.
Numerically, the bound is
$\rho_{\mu e}\rho_{tt}\lesssim 0.4\, (0.5)\, \l_e \l_t$ %($0.5 \l_e\l_t$)
for $\Delta m = 40$ (200) GeV. %, respectively.

%\tcr
{Since our focus is primarily on implications from NP in muon $g-2$,
we do not discuss combined result of one- and two-loop effects.
The latter not only involves couplings inconsequential to
muon $g-2$, it also tends to cancel against the one-loop contribution, 
which will only bring uncertainty into our predictions of $\mtoeg$ 
given in Fig.~\ref{fig:mufv}.
One also needs to think about the complex phase of $\rtt$, 
which is of interest for EWBG.}

%TODO $\mu\to 3e$
{\bm{$\mu\to eee$}.---} \ In g2HDM, neutral scalar exchange 
with couplings $\rho_{\mu e}$ and $\rho_{ee}$ can induce $\mu \to 3e$,
and the expression for the branching ratio
can be found in Ref.~\cite{Hou:2020itz}. 
Eq.~\eqref{eq: l_1j} then implies that the contribution is very small~\cite{Hou:2020itz}.
For example, with $m_{H\, (A)} = 300$ (340)~GeV %and $m_A = 340$~GeV, 
we find ${\cal B}(\mu \to 3e)\simeq 3 \times 10^{-24}$, 
and even more suppressed for larger $m_A$.  
This is far below the SINDRUM bound~\cite{SINDRUM:1987nra} 
of ${\cal B}(\mu \to 3e) < 10^{-12}$.
The Mu3e experiment plans to push the limit 
down to $10^{-16}$~\cite{Blondel:2013ia}, which 
falls short by many orders of magnitude.
%means that 
%probing for electron-related extra Yukawa coupling via $\mu \to 3 e$ is not likely. 

However, the $\mu e \gamma $ dipole can  generate sufficiently
large contribution to $\mu \to 3 e$ via~\cite{Kuno:1999jp}
\begin{align}
 {\cal B}(\mu \to 3e) \simeq \frac{\alpha}{3\pi}
    \left[{\rm log}\left(\frac{m_\mu^2}{m_{e}^2}\right) - \frac{11}{4} \right]
 {\cal B}(\mu \to e\gamma).
 \label{eq:mu3e}
\end{align}
We plot in Fig.~\ref{fig:mufv} the parameter space (blue shaded) 
probed by Mu3e, assuming Eq.~\eqref{eq:mu3e}. 
Note that the upper boundary does not correspond to the SINDRUM bound, 
as it is already ruled out by MEG. Furthermore,
we show only the parameter space that lies beyond 
the reach of MEG~II. %, but within Mu3e reach. 
This implies an important aspect of g2HDM: 
a discovery by MEG~II means that Mu3e would 
observe $\mu\to 3 e$ and confirm the dipole behavior.
%More optimistic for Mu3e, 
If MEG~II does not see any hint, Mu3e can probe the dipole coupling further
and find hints if $\rte \gtrsim  \l_e/3$ for near-degeneracy $\Delta m = 40$ GeV case. 
For larger $\Delta m $, Mu3e can access smaller values of $\rte < \l_e/3$ 
(Fig.~\ref{fig:mufv}(right)).

\section{\bm{$\mu  \to \lowercase{e}$} Conversion on Nuclei}

Concerning experimental prospects for muon flavor violation,
it is the $\mu N \to e N$ process where the ultimate progress would occur.
The experimental bound is quoted in terms of the ratio, $R_{\mu e}$,
which is defined as the $\mu\to e$ conversion rate normalized to the
muon capture rate~\cite{Kuno:1999jp},
\begin{equation}
	R_{\mu e} = \frac{\Gamma(\mu +(A, Z) \to e + (A, Z))}
                                {\Gamma(\mu + (A, Z) \to \nu_\mu + (A, Z-1))},
\end{equation}
with $A$ and $Z$ the mass and atomic numbers of the target nucleus, respectively.
SINDRUM~II gave the bound of 
$R_{\mu e} < 7 \times 10^{-13}$~\cite{SINDRUMII:2006dvw} 
using gold as target. 

An array of experiments aim to improve the sensitivity in the near future.
DeeMe~\cite{Teshima:2018ise} plans to reach a
sensitivity of $10^{-14}$ using thick silicon carbide (SiC) target.
COMET \cite{COMET:2018auw} and Mu2e~\cite{Mu2e:2014fns}
aim at reaching $\sim 10^{-17}$ using aluminum (Al) target, 
while PRISM~\cite{Kuno:2005mm} can push the sensitivity to $10^{-18}$
with titanium (Ti) target.
The experimental prospects seem quite promising.

In the context of g2HDM, the relevant effective Lagrangian for
$\mu \to e$ conversion is given by Eq.~\eqref{eq:Leff-dipole} 
plus Fermi contact terms~\cite{Cirigliano:2009bz,Harnik:2012pb,Crivellin:2014cta},
\begin{equation}\label{eq: mu2eN}
	\mathcal{L}_{\mathrm{eff}}^{\mu \to e} =
	\mathcal{L}_{\mathrm{eff}}^{\rm \mu e\gamma}
	+  {\sum_q} \bigl(C_{q q}^{S R}\, \bar{e} L \mu
               + C_{q q}^{S L}\, \bar{e} R \mu\bigr)\, m_{\mu} m_{q} \bar{q} q, 
\end{equation}
where $C_{q q}^{S R(L)}$ arise from neutral scalar exchange~\cite{Hou:2020itz}.
{Note that the  Wilson coefficients in Eq.~\eqref{eq: mu2eN} are
by definition invariant under one-loop QCD renormalization.\footnote{We ignore
running due to QED in view of $\alpha_e \ll \alpha_s$. The analysis of such effects
in effective theory approach can be found in
Refs.~\cite{Czarnecki:2001vf,Pruna:2014asa,Davidson:2016edt,Crivellin:2017rmk}}
Therefore, the values of running quark masses and couplings $\rho_{qq}$, which enter
 the expressions of $C_{q q}^{S R(L)}$ (given in Ref.~\cite{Hou:2020itz}), should be
taken at the same scale.}
The conversion rate, $\Gamma_{\mu \to e}$, is then defined as,
\begin{equation} \label{eq: Gam_mue}
\begin{aligned}
\Gamma_{\mu \rightarrow e} = {m_{\mu}^{5}}
  \left| {1\over 2} C_{T}^{L} D +
               2 \sum_q\Bigl(m_{\mu} m_{p}\,  C_{qq}^{SL}f_q^p S^{p}\right.\\
      + \left. p \rightarrow n\Bigr)\right|^{2} + (L \to R),
\end{aligned}
\end{equation}
where the coefficients $D$ and $S^{p(n)}$ are related to lepton-nucleus
overlap integrals, and $f_q^{p(n)}$ are nucleon matrix elements.
For gold nuclei, $D = 0.189$, $S^p = 0.0614$, $S^n = 0.0918$~\cite{Kitano:2002mt},
while $D=0.0362$, $S^p=0.0155$, and $S^n = 0.0167$ \cite{Kitano:2002mt} 
for aluminum. The values of $f^{p(n)}$are taken 
from Ref.~\cite{Harnik:2012pb} for $u$ and $d$ quarks,
from Ref.~\cite{Junnarkar:2013ac} for the $s$ quark, and we use 
the relation~\cite{Shifman:1978zn}
$f_Q^{p(n)}= (2/27)(1-f_u^{p(n)}-f_d^{p(n)}-f_s^{p(n)})$
for the heavy quarks $c, b, t$.

{\it \textbf{Dipole dominance}.---} \
In the absence of extra quark {Yukawa} couplings $\rho_{qq}$,
the conversion rate in Eq.~\eqref{eq: Gam_mue} is governed
by the $\mu e \gamma$ dipole.
In the dipole dominance scenario,  Eq.~\eqref{eq: Gam_mue}
can be written in terms of the $\mtoeg$ decay rate in a model independent way.
For a given target nuclei with overlap integral coefficient $D$, one finds,
\begin{equation}\label{eq: mu2e-dipole}
	\Gamma_{\mu \to e} \simeq \pi D^2\,\Gamma(\mtoeg).
\end{equation}
With the knowledge of the muon capture rate for a given nuclei, 
one can estimate the conversion ratio, $R_{\mu e}$. 
For Au and Al, the muon capture rates are 
$13.07 \times 10^6 \,s^{-1}$ and $0.71 \times 10^{6}\, s^{-1}$,
respectively~\cite{Kitano:2002mt,Suzuki:1987jf}.
For other nuclei, the muon capture rates 
can be found in Ref.~\cite{Kitano:2002mt}.

Taking Al as target nuclei, we illustrate in Fig.~\ref{fig:mufv} 
the region (gray shaded) where the upcoming $\mu N\to eN$ experiments 
will make further improvements in probing the g2HDM parameter space. 
The lower boundary shows the experimental sensitivity of $10^{-17}$. 
It is no surprise that $\mntoen$ will be probing the $\mu e \gamma$ dipole 
the furthest among all three processes, 
given the projected vast improvements in sensitivity.
Even if $\rte$ turns out to be
an order of magnitude smaller than our conservative suggestion
of $\rte \simeq {\cal O}(\l_e)$, a discovery %in $\mntoen$ 
is still feasible.

Fig.~\ref{fig:mufv} also shows the possible interplay of 
different experiments. %on muon flavor violation. 
If MEG~II finds hint of $\mtoeg$, then Mu3e and
Mu2e/COMET can confirm the dipole-only behavior, in accord 
with Eqs.~\eqref{eq:mu3e} and \eqref{eq: mu2e-dipole}. %, respectively. 
However, if MEG~II does not see any hint,
there is still window (blue shaded region) for Mu3e discovery,
which again would likely be dipole-induced in g2HDM, 
and can help interpret any Mu2e/COMET confirmation. 
A discovery solely at Mu2e/COMET, however, leaves room 
for speculation on the nature of the interaction responsible for 
the hint --- dipole-like or contact scalar interactions.

\begin{figure*}[t]
\center
\includegraphics[width=.35\textwidth]{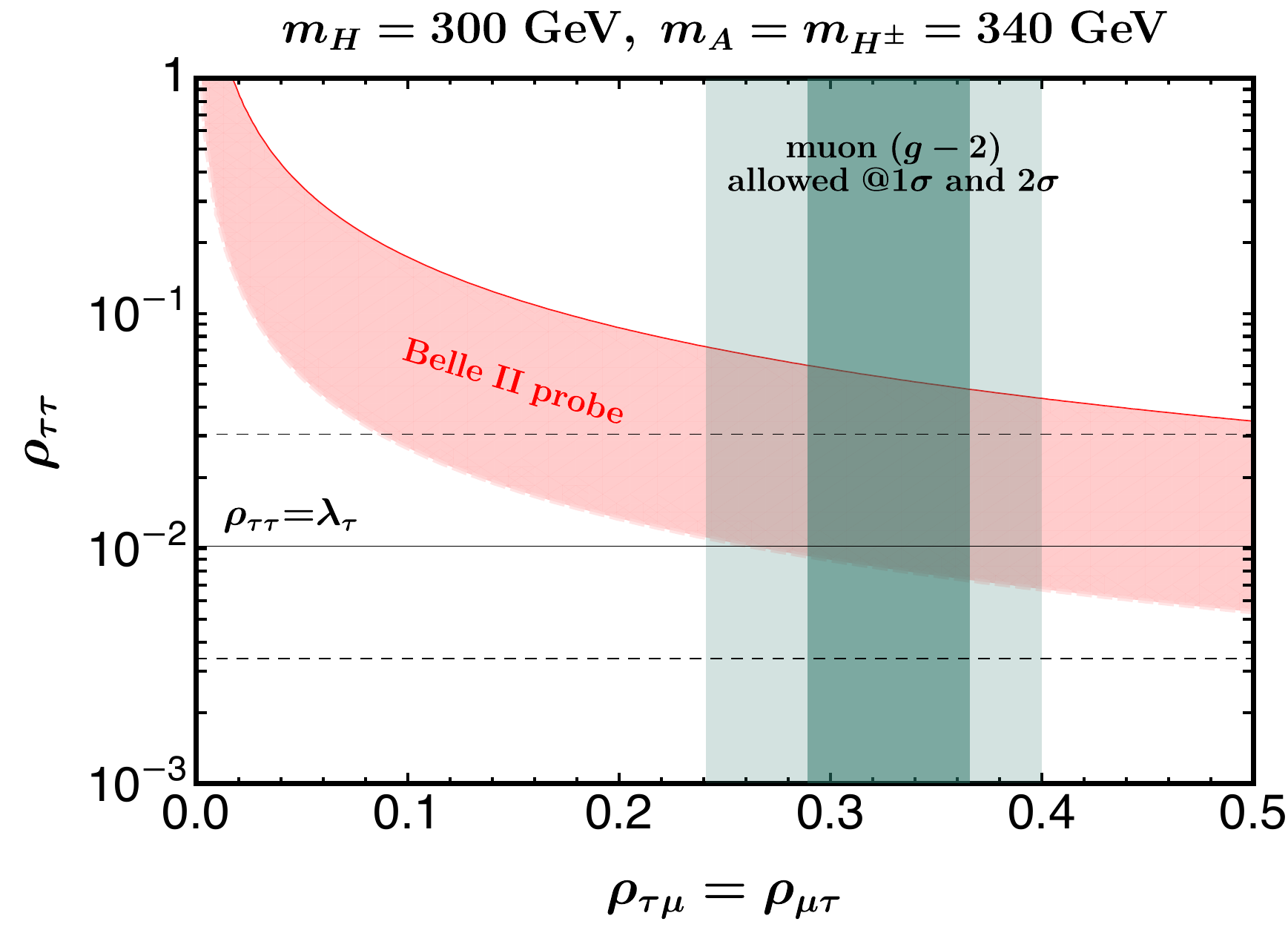}
\hskip 0.3cm
\includegraphics[width=.35\textwidth]{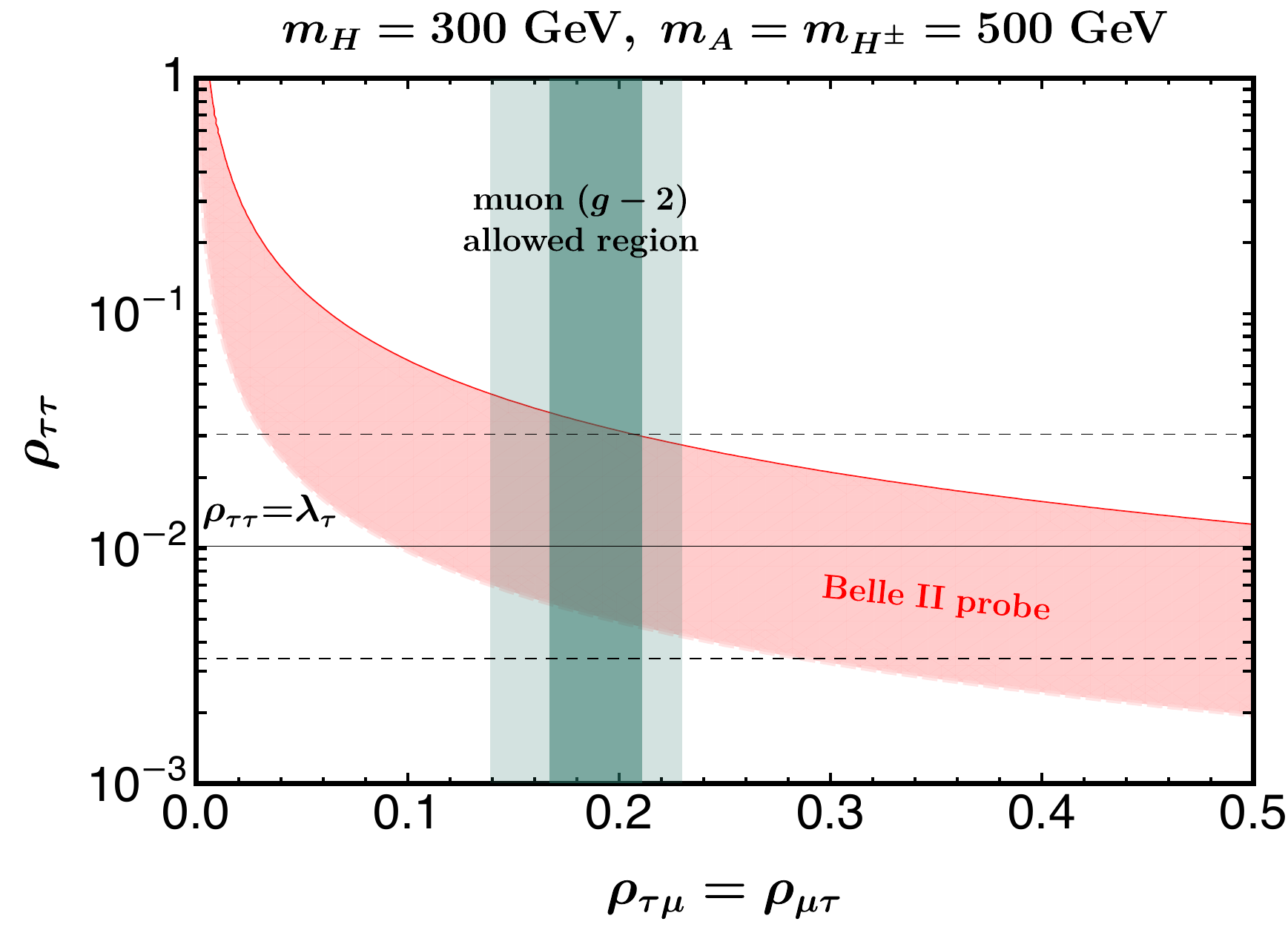}
\caption{
Parameter space in the $\rtm$--$\rtata$ plane 
for $\tau \to \mu \gamma$ to be probed by Belle~II.  
See text for details.}
\label{fig:taufv}
\end{figure*}

{\it \textbf{Contact interactions}.---} \
The $\mntoen$ process provides a distinct probe of charged LFV 
compared with $\mtoeg$, because of scalar interactions in 
Eq.~\eqref{eq: mu2eN} that could be significant,
due to diagonal extra $\rho_{qq}$ couplings.  
%Unlike $\mtoeg$, the 
Tree diagrams for $\mu \to e$ conversion
do not exhibit cancellation between $H$ and $A$ contributions. 
In fact, for real $\rho_{ij}$, $A$ does not contribute to coherent $\mntoen$.
This unique feature of $\mntoen$ conversion can be 
exploited to probe the mass hierarchy of $H$ and $A$, 
in the context of one-loop solution to muon $g-2$ in g2HDM. 

As mentioned in Sec.~\ref{sec: II}, muon $g-2$ admits both 
$m_A-m_H>0$ and $m_A-m_H<0$, 
because one has freedom in the sign of the $\rtm \rmt$ product
to contribute positively to $\Delta a _\mu$.\footnote{We adopt $\rmt \rtm>0$,
which is compatible with 
our assumption of real couplings $\rho_{ij}=\rho_{ji}$. 
For $\rmt \rtm<0$, Ref.~\cite{Omura:2015xcg} used $\rtm =-\rmt$.} 
Since $\mntoen$ depends on the mass of $H$ only (in case of real coupling), 
a lighter $H$ will contribute significantly more.
For example, taking $\rho_{\mu e} = \l_e$, $\rho_{qq} = \l_q$
 {for $q\neq t$, and $\rho_{tt} \sim 0.1$ (0), with Al as target we obtain 
$R_{\mu e}|^{\rm contact} \simeq 1\, (0.95) \times 10^{-16}$} for $m_H = 300$~GeV.
Since $R_{\mu e}|^{\rm contact}$ scales as $(1/m_H^4)$, 
it is quickly damped for heavier $H$.
The tree level prediction in g2HDM is somewhat uncertain, 
given the large number of couplings involved. 
However,  our estimate of ${\cal O}(10^{-16})$, 
which lies well within experimental reach, is still a conservative estimate, 
as can be seen from our choice of  $\rho_{qq}$ values.
This brings up another interesting aspect of $\mntoen$, 
which is the interference of dipole and contact interactions. 
If both contributions have comparable strength,
constructive interference  --- the most optimistic case --- 
can catapult the value of $R_{\mu e}$ to be much larger
than $10^{-16}$. 
% enough  for even a concurrent discovery with 
% a MEG~II observation of the $\mu e \gamma$ dipole! 
%
On the other hand, if $A$ is the lighter one while $H$ is heavy,
the contact effect could be quite suppressed.

With results of MEG~II, Mu3e, and COMET/Mu2e 
playing out and unfolding in the next decade, 
probing for flavor violation in muon decays look promising, 
{\it if} muon $g-2$ is due to a large $\rho_{\mu\tau}$ coupling.

\section{\bm{ $\ttomg$} }

%In g2HDM, $\ttomg$ is actually more closely correlated to muon $g-2$.
%The one-loop diagram\footnote{Again, loops involving lighter leptons are
%chirally suppressed and neglected.} are similar
%to Fig.~\ref{fig: one-loop} where one just replaces labels $``\mu" \to ``\tau"$
%with $\ell=\mu$ . 
%both can be induced with single coupling $\rtm$---together with a nonzero
%value of mixing angle $c_\gamma$. However, emergence of alignment
%phenomena ($\cg \to  0 $) at LHC implies a vanishing $\ttomg$;
%and especially in the context of $\tau-\mu$ flavor violation the
%CMS search for $h\to\tau\mu$ provide stringent constraint on
%the product $\rtm c_\gamma\lesssim 0.1 \l_\tau$, making $\bttomg$
%fall below the sensitivity range which will be probed in
%next couple of years. But, other extra Yukawa couplings, and especially the diagonal
%ones, $\rtata$ (and $\rtt$ \cite{Hou:2020tgl, Hou:2020itz}) can easily elevate
%the $\bttomg$ to within experimental reach.

The Belle experiment updated recently the bound on $\ttomg$, 
giving $\bttomg_{\rm Belle} < 4.2 \times 10^{-8}$~\cite{Belle:2021ysv}. 
The projected limit by Belle~II 
%expects to improve by more than an order of magnitude, to 
is to push down to $10^{-9}$~\cite{Belle-II:2018jsg}, 
and discovery is possible.

The physics of $\ttomg$ in g2HDM is similar to
$\mtoeg$ discussed in Sec.~\ref{sec: II}. 
One just replaces the label $``\mu" $ with $``\tau" $ and set $\ell=\mu$
for the one-loop diagram of Fig~\ref{fig: one-loop}, 
and corresponding expressions can be obtained from Eq.~\eqref{eq:1-loop}. 
Again, loops involving lighter leptons are chirally suppressed and neglected. 
We plot in Fig.~\ref{fig:taufv} the region (red shaded) in
the $\rtm$--$\rtata$ plane probed by Belle~II in g2HDM.
The upper boundary is the Belle bound~\cite{Belle:2021ysv}, 
while the lower boundary is the Belle~II projection~\cite{Belle-II:2018jsg}.   
With large $\rtm$ giving $1\sigma$ solution to muon $g-2$, 
the Belle limit already probes $\rtata\simeq 6 \l_\tau$ 
for near-degenerate scalar masses (left plot). 
Belle~II will continue to probe lower values and can 
push down to $\rtata\simeq \l_\tau$ with full data. 
For large $m_H-m_A$ mass difference (right plot), 
hence larger one-loop contribution, Belle~II can probe smaller $\rtata$ values.
The Belle bound would already rule out  $\rtata > 3\l_\tau$
in regions allowed by muon $g-2$, while Belle~II can probe
below $\rtata \sim \l_\tau$.

It is interesting that one does not require a large value of 
extra $\rho_{\tau\tau}$ coupling for $\ttomg$ to be discoverable at Belle~II.
Analogous to the $\mu \to e\gamma$ discussion, 
a {\it natural} $\rtata\simeq {\cal O}(\l_\tau)$ would suffice,
so long that the one-loop mechanism is behind the muon $g-2$ anomaly.

%TODO $ttomg$ 2-loop
{\it \textbf{Two-loop mechanism}.---} \
In contrast to the $\mtoeg$ case, since
the $\rtm$ coupling enters the Barr-Zee diagrams for $\ttomg$ directly, 
the two-loop contributions have far more significant implications
for extra top {Yukawa} coupling~\cite{Hou:2020tgl, Hou:2020itz}, and for 
realizing EWBG in g2HDM~\cite{Fuyuto:2017ewj,Fuyuto:2019svr,Hou:2020tnc}.
We have discussed these contributions and the implications
for muon $g-2$ and at the LHC in Ref.~\cite{Hou:2021sfl}. 
For completeness, let us give a quick recount of the results. 

For $m_H,\, m_A =300$, $340$ GeV, the dominant
two-loop diagrams involving top give the strict bound of $\rtt\lesssim 0.05$.
For heavier pseudoscalar, $m_A=500$ GeV, the bound gets slightly
relaxed, $\rtt\lesssim 0.1$. 
However,  it turns out that the LHC search for $gg \to H,A \to \tau\mu$ 
provides a stronger constraint~\cite{Hou:2021sfl} than $\ttomg$.
Now, having $\rtt\gtrsim 0.1 $ would make a more robust 
driver for EWBG in g2HDM \cite{Fuyuto:2017ewj}.
One could then bring in another extra top {Yukawa} coupling, $\rho_{tc}$, 
to dilute $H, A\to \tau\mu$ and relax~\cite{Hou:2021sfl} the constraint on $\rtt$. 
If $\rtt$ is indeed very small in g2HDM, 
then $\rho_{tc}$ can play the role of the EWBG driver~\cite{Fuyuto:2017ewj}.
Although it has nothing to do with muon $g-2$, it could lead to interesting signals
such as $cg \to bH^+ \to \tau^\pm \mu^\mp b W^+, t\bar cbW^+$
at the LHC~\cite{Hou:2021sfl}.   

For real and positive values of extra {Yukawa} couplings, there is 
cancellation between one- and two-loop contributions,
since the top loop of the latter brings in an extra minus sign. 
With $\rtm$ fixed from one-loop solution to muon $g-2$, the cancellation
can enlarge the $\rtata$ and $\rtt$ parameter range,
%that are consistent with $\ttomg$, 
as pointed out in Ref.~\cite{Omura:2015xcg}.
But this fine-tuned parameter space is not quite likely to survive.
One reason could be the complex nature of $\rtt$ 
that is needed for EWBG, making the two-loop amplitude complex, 
and e.g. for phase of $\rtt$ at $\pm \pi/2$ would make the cancellation mute.
A second reason is experimental: the parameter space with 
$\rtm$, $\rtata$, and $\rtt$ simultaneously large is actually
under stress from LHC searches~\cite{Hou:2021sfl}.

%TODO $\tau\to 3\mu$
{\it \bm{$\tau\to \mu\mu\mu$}.---} \
This decay has been, and will be, searched for by several experiments. 
The current Belle limit~\cite{Hayasaka:2010np} is 
${\cal B}(\tau \to 3\mu)|_{\rm Belle}< 2.1 \times 10^{-8}$.
Both Belle~II~\cite{Belle-II:2018jsg} and
LHCb (Upgrade~II)~\cite{LHCb:2018roe} plan to improve the sensitivity,
with Belle~II projecting a better reach of $3.3 \times 10^{-10}$.
There is also a new fixed-target proposal,
TauFV \cite{Beacham:2019nyx}, aiming to reach $\sim 10^{-10}$ sensitivity.
So the experimental prospect looks good.

Similar to $\mu\to 3e$, this decay is induced at 
tree-level by $H$, $A$ FCNC in g2HDM~\cite{Hou:2020itz}. 
But unlike $\mu\to 3e$, tree-level $\tau\to 3\mu$ 
involves not only large $\rtm$, but also the $\rho_{\mu\mu}$ coupling, 
and ${\cal B}(\tau\to3 \mu)$ can be significant. 
For our benchmark cases of $\Delta m = 40$ and $200$ GeV,
we find the current experimental bound only constrains
$\rtm \rho_{\mu\mu} \lesssim 260 \l_\tau \l_\mu$ and
$320 \l_\tau \l_\mu$, respectively. 
Taking the future sensitivity of Belle~II, we get
$\rtm \rho_{\mu\mu} \lesssim 32 \l_\tau \l_\mu$ and $40 \l_\tau \l_\mu$, respectively. 
This means that, for $\rtm\simeq 30 \l_\tau$  ($20\l_\tau$) 
for $\Delta m =40$  (200) GeV, %to explain the muon $g-2$ anomaly,
discovery is projected in g2HDM even with $\rho_{\mu\mu} \simeq {\cal O}(\l_\mu)$,
which is again a ``natural'' strength in g2HDM.

If $\rho_{\mu\mu} \ll \l_\mu$ turns out to be the case in {\it Nature}, 
then $\tau\to3\mu$ search will essentially be probing the $\tau \mu \gamma$ dipole, 
which relates to $\tau\to 3\mu$ by changing 
``$\mu$'' to ``$\tau$'' and ``$e$'' to ``$\mu$''in Eq.~\eqref{eq:mu3e}.
This means that  ${\cal B}(\tau\to 3\mu)$ will be below
${\cal B}(\tau\to\mu\gamma)$ by about $2.3 \times 10^{-3}$.
Therefore, as argued in Ref.~\cite{Hou:2020itz}, 
unless there is a hint of $\tau\to\mu\gamma$ in the early data of  Belle~II,
${\cal B}(\tau\to3\mu)$ will be outside the sensitivity reach of planned experiments.

\section{Discussion and Summary}

We find that $\mtoeg$ can be enhanced in g2HDM to 
experimentally accessible values, even for exceptionally small 
extra {Yukawa} coupling $\rte = {\cal O}(\l_e)$. 
This is in context of using large $\rtm$ coupling to explain the muon $g-2$ anomaly.
A diagram similar to Fig.~\ref{fig: one-loop} with $\tau$ and 
$H,\,A$ in the loop can contribute to {electron $g-2$}{,
where recent measurements of $\alpha$ suggest some tension~\cite{Parker:2018vye}.
} 
But since large $\rtm$ constrains $\rte$ ($\ret$), 
through the MEG bound on $\mu \to e\gamma$, 
to be consistent with Eq.~\eqref{eq: l_1j}, 
we find the contribution is negligible and the electron  $g-2$ remains SM-like. 
The $\rte$, $\ret$ couplings, together with $\rtata$, 
induce $\tau\to e\gamma$ decay. 
But again with Eq.~\eqref{eq: l_1j} and with $\rtata = {\cal O}(\l_\tau$),
the induced {${\cal B}(\tau\to e\gamma)$ is very small.}
Putting it differently, the current bound of
${\cal B}(\tau\to e\gamma) < 3.3 \times 10^{-8}$~\cite{BaBar:2009hkt} sets
only an extremely poor bound of $\rte \rtata \lesssim {\cal O}(10^4) \lambda_e \lambda_\tau$
for scalar masses considered in this work, and far from probing Eq.~\eqref{eq: l_1j}.

We mention some similarly weak constraints in passing. 
The $Z\to\tau\tau$, $\mu\mu$ partial widths and the leptonic
$\tau \to \mu \nu \bar\nu$, $e \nu \bar\nu$ and $\mu \to e \nu \bar \nu$ decays
are easily compatible with Eq.~\eqref{eq: l_1j}
and $\rho_{\tau\tau} \simeq {\cal O}(\l_\tau)$~\cite{Hou:2021sfl}
and even large $\rho_{\tau\mu}$,
and no serious constraint is set within our framework.

{
Because of the Fermilab confirmation of the muon $g-2$ anomaly, 
we have taken the one-loop explanation in g2HDM seriously. 
We had not advocated this in our previous work~\cite{Hou:2020itz}, 
but it should be clear that {\it Nature} is entitled to this 
choice of a large $\rho_{\tau\mu}\, (\simeq \rho_{\mu\tau})$, 
which has phenomenological consequences such as small $\rho_{tt}$ 
and $\rho_{\tau e} = {\cal O}(\lambda_e)$.
We would still not advocate that {\it Nature} can whimsically
dial up {\it several} extra {Yukawa} couplings, for it would seem
hard to escape the exquisite flavor probes.
Another curiosity worth emphasizing again is that, if it turns out 
that $A$ is the lighter exotic scalar behind muon $g-2$
and $H$ is considerably heavier, since $A$ exchange cannot be
coherent over the nucleus, the contact interaction effects
could be much subdued, hampering the $\mu N \to eN$ program
to study them.
}

{
In summary, in the general two Higgs doublet model, 
large LFV couplings $\rho_{\tau\mu}$ and $\rho_{\mu\tau}$, 
with inbuilt chiral enhancement, can explain the muon $g-2$ anomaly. 
We cover LFV processes such as $\mu\to e\gamma,\,  3e$,  $\mntoen$, $\ttomg,\, 3\mu$. 
Taking $\rho_{\tau e},\, \rho_{\mu e} = {\cal O}(\lambda_e)$ (Eq.~\eqref{eq: l_1j}) 
and $\rho_{\ell\ell} \simeq {\cal O}(\l_\ell)$ as reasonable,
we find excellent chance for discovery of $\mu e\gamma$ dipole effects with 
all three muon decay/transition experiments.
Among these, $\mntoen$ conversion can ultimately determine 
or constrain the associated LFV couplings.
If extra quark {Yukawa} couplings come into play,
$\mntoen$ can probe the interference between dipole and contact interactions. 
By exploiting different nuclei and refined
theory developments, the $\rho_{qq}$ couplings might be unraveled. 
Prospects for $\tau\to \mu \gamma$ and $\tau\to 3\mu$ are also good, 
which probe the natural strengths of $\rho_{\tau\tau} \sim \lambda_\tau$
and $\rho_{\mu\mu} \sim \lambda_\mu$.

Let us hope that the muon $g-2$ anomaly would
usher in a new era of $\mu/\tau$ discoveries.
}

\vskip0.2cm
\noindent{\bf Acknowledgments} \
This research is supported by MOST 109-2112-M-002-015-MY3 
and 109-2811-M-002-540 of Taiwan, and NTU 110L104019 and 110L892101.

%-----------------------------------------------------------------------------------------------------------------------------------

\end{document}